\documentclass[journal]{IEEEtran}
\usepackage{graphicx}
\usepackage{graphics}

\begin{document}
\title{Research and Development of Commercially Manufactured Large GEM Foils}
%
%

\author{M.~Posik and B.~Surrow%

\thanks{Manuscript received November 26, 2014. This work was supported in part by an EIC R\&D subcontract \#223228, managed by Brookhaven National Laboratory and The College of Science and Technology at Temple University.}
\thanks{M.~Posik is with Temple University, Philadelphia, PA 19122 USA (e-mail: posik@temple.edu).}%
\thanks{B.~Surrow is with Temple University, Philadelphia, PA 19122 USA (e-mail: surrow@temple.edu).}%
}

\maketitle
\pagestyle{empty}
\thispagestyle{empty}

\begin{abstract}
The recently completed Forward GEM Tracker (FGT) of the STAR experiment at RHIC took advantage of commercially produced GEM foils based on double-mask chemical etching techniques. With future experiments proposing detectors that utilize very large-area GEM foils, there is a need for commercially available GEM foils. Double-mask
etching techniques pose a clear limitation in the maximum size. In contrast, single-mask techniques developed at CERN would allow one to overcome those limitations.
We report on results obtained using 10 $\times$ 10 cm$^2$ and 40$\times$40 cm$^2$ GEM foils produced by Tech-Etch Inc. of Plymouth, MA, USA using single-mask techniques and thus the beginning for large GEM foil production on a commercial basis. A quality assurance procedure has been established through electrical and optical analyses via leakage current measurements and an automated high-resolution CCD scanner. The Tech-Etch foils show excellent electrical properties with leakage currents typically measured below 1 nA. The geometrical properties of the Tech-Etch single-mask foils were found to be consistent with one another, and were in line with geometrical specifications from previously measured double-mask foils. The single-mask foils displayed good inner and outer hole diameter uniformities over the entire active area.
\end{abstract}


\section{Introduction}
%
%
%
%
\IEEEPARstart{T}{echnology} based on gas electron multipliers (GEMs) have been establishing their presence in the 
nuclear and particle physics communities since their invention in 1997~\cite{Sauli:1997qp}. They have several attractive features including 
the ability to perform in a high rate environment ($>10^5$ Hz/mm$^{2}$~\cite{Altunbas:2002ds}), excellent spatial resolution (40 $\mu m$ rms~\cite{Altunbas:2002ds}), and the ability to cover a large acceptance. Several experiments: STAR~\cite{Surrow:2010zza}, COMPASS~\cite{Altunbas:2002ds}, and others are already employing the use of GEM technology in their detectors. With GEM technology maturing and based on successful runs from
experiments already using GEM technology, many future experiments and experiment upgrades are either planning on or looking into using GEM technology, such as ALICE~\cite{Gasik:2014sga}, JLab's Super BigBite Spectrometer~\cite{SBS}, CMS~\cite{Abbaneo:2014lja} and dedicated EIC detectors. 

The main distributor of GEMs to the scientific community is CERN. In the past CERN has been able to adequately provide GEMs to experiments that needed them. However, given the newly generated interest in GEMs and the fact that CERN is not a dedicated production facility, one can not expect CERN to be able to provide all experiments with the GEMs that they need. As a result the commercialization of GEMs has been successfully established at Tech-Etch Inc~\cite{TechEtch}, which will help to alleviate the high demand for GEMs. Tech-Etch Inc. is a company based in Plymouth, Massachusetts who have commercialized large area (up to $\sim$50$\times$50 cm$^{2}$) GEMs using single-mask and double-mask etching processes~\cite{Surrow:2010zza,Becker:2006,Surrow:2007em,Simon:2008bv}.          

\section{Single-Mask Etching Process}
\begin{figure}
\centering
\includegraphics[width=3.5in]{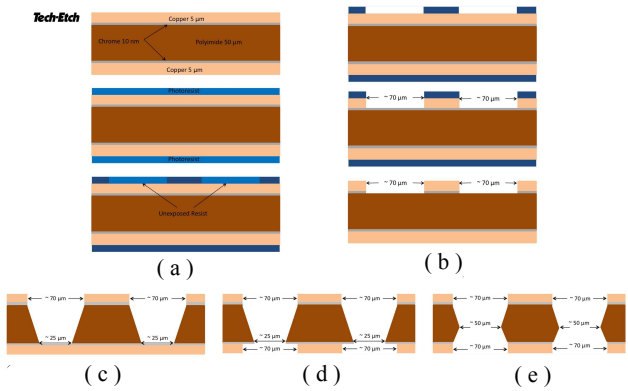}
\caption{Overview of the GEM production using the single-mask process at Tech-Etch. See text for description.}
\label{fig:single-mask-process}
\end{figure}

\begin{figure}[!t]
\centering
\includegraphics[width=0.5\columnwidth, angle=-90]{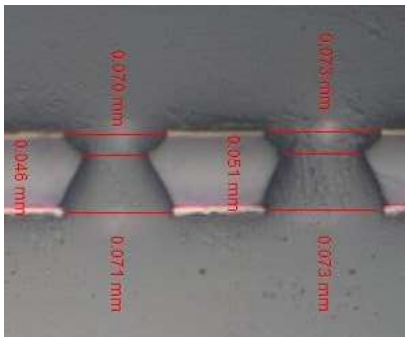}
\caption{Cross section image of single-mask 40$\times$40 cm$^2$ Tech-Etch produced GEM foil.}
\label{fig:gem-cross-section}
\end{figure}

 Tech-Etch employs the single-mask etching technique to produce their GEM foils. This has the advantage of allowing for the production of larger GEMs, up to about 1 m long. Figure~\ref{fig:single-mask-process} highlights the GEM foil production steps used by Tech-Etch to produce their single-mask GEMs. A GEM is produced starting from a standard foil which has a polyimide layer made of Apical, that is about 50 $\mu m$ thick, which is sandwiched between two layers of copper ($\sim$ 5 $\mu m$ thick). The foil is then coated with a layer of photoresist, and laser direct imaging is used to apply the micro hole pattern to the front side of the foil (fig.~\ref{fig:single-mask-process} (a)). The unexposed photoresist is developed away and the front side copper is etched via an acid bath (fig.~\ref{fig:single-mask-process} (b)). Ethylenediamine (EDA) chemistry is then used to etch the polyimide layer on the front side of the foil (fig.~\ref{fig:single-mask-process} (c)) and the back side copper layer is then etched using an electrolytic process (fig.~\ref{fig:single-mask-process} (d)). This now leaves the micro pattern holes on the GEM foil having a conical structure. In order to obtain the desired double conical hole geometry, the back side polyimide layer is etched using EDA chemistry (fig.~\ref{fig:single-mask-process} (e)). The resulting double conical structure can clearly be seen in the cross-sectional image of a Tech-Etch 40$\times$40 cm$^2$ GEM foil shown in fig.~\ref{fig:gem-cross-section}.

\section{Tech-Etch GEM Production}
Following the GEM production process discussed in the previous section, Tech-Etch has successfully produced 10$\times$10 and 40$\times$40 cm$^2$ GEM foils. Three 10$\times$10 cm$^2$ manufacturing lots consisting of 6,12, and 6 foils respectively, have been sent to Temple University for analysis of their electrical performance and geometrical properties. Additionally one 40$\times$ 40 cm$^2$ manufacturing lot consisting of 3 foils was also sent to Temple University for analysis. Section~\ref{sec:measurements-techniques} will discuss the means by which the electrical performance and geometrical properties of the foils were determined. Sections~\ref{sec:electrical} and~\ref{sec:geometrical} will present the electrical and geometrical results, respectively.
 
\section{Measurement Techniques}\label{sec:measurements-techniques}
\begin{figure}[!t]
\centering
\includegraphics[width=0.75\columnwidth]{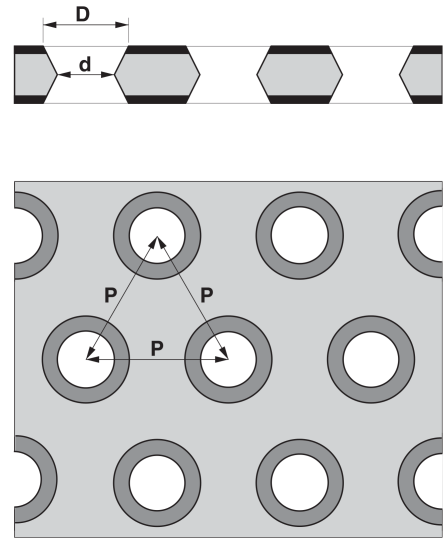}
\caption{Schematic view of Tech-Etch single-mask GEM foil. Image reproduced from ref~\cite{Becker:2006}.}
\label{fig:gem-geometry}
\end{figure}

The production quality a GEM foil can be quantified through its electrical and geometrical properties. The electrical performance of the GEM is determined through its leakage current. This is measured by applying a high voltage across the GEM foil and measuring the resulting current. The geometrical quality is determined through an optical analysis where the pitch (P) between two neighboring holes, the inner hole diameter (d, determined from the polyimide layer) and the outer hole diameter (D, determined from the copper layer) are measured, as shown in fig.~\ref{fig:gem-geometry}.

\subsection{Electrical Analysis}\label{sec:electric-ana}
\begin{figure}[!t]
\centering
\includegraphics[width=\columnwidth]{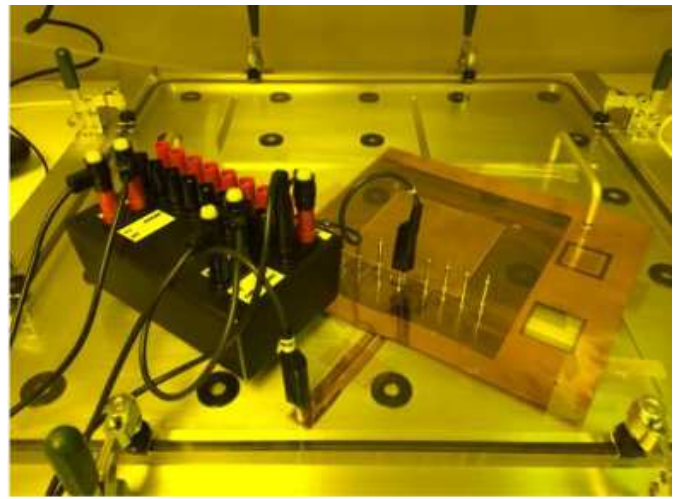}
\caption{Electrical testing setup with a 10$\times$10 cm$^2$ GEM foil enclosed in a nitrogen box ready for leakage current measurements.}
\label{fig:electric-setup}
\end{figure}

The leakage current measurements were performed in a class 1,000 clean room at Temple University. Within the clean room the GEM foil was placed into a Plexiglas enclosure, as shown in fig.~\ref{fig:electric-setup} which was continuously flushed with nitrogen to help prevent sparking and any debris from settling on the foil. After at least an hour of flushing nitrogen gas through the enclosure, a voltage was applied to the GEM foil and slowly ramped from 0 to 600 V, where the leakage current was measured in 100 V increments. The voltage was applied across the foil and its current measured using an ISEG SHQ 222M high voltage power supply.

\subsection{Optical Analysis}\label{sec:optical-ana}
\begin{figure}[!t]
\centering
\includegraphics[width=\columnwidth]{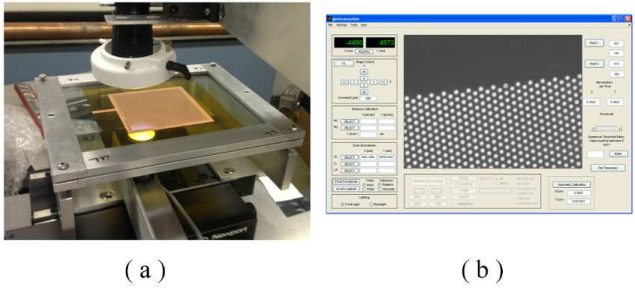}
\caption{(a) Optical scanner setup for 10$\times$10 cm$^2$ GEM foils. (b): Image of inner hole diameters on MATLAB software.}
\label{fig:ccd-setup}
\end{figure}

The geometrical properties of the foil were measured using an automated 2D CCD scanner (fig.~\ref{fig:ccd-setup} (a)). The setup used at Temple University is identical to that which is described in ref.~\cite{Becker:2006}. The CCD camera setup consists of a video camera connected to a 12x zoom lens through a 2x adapter, with a ring of LEDs around the lens face (front light). The CCD setup is coupled to a support stage with an LED light mounted below it (back light). The stage is able to traverse in two dimensions which allows the entire active area of the GEM foil to be scanned with high precision. The GEM foil is enclosed between two glass plates which are secured by an aluminum frame. The apparatus is controlled through a MATLAB graphical interface (fig.~\ref{fig:ccd-setup} (b)). The sensitivity of the CCD camera to the GEM's inner or outer hole diameters is determined by the lighting scheme used. If the front light is used to illuminate the GEM, then we are sensitive to the GEM's outer hole diameters. On the other hand, if the back light is used to illuminate the GEM, then the measurements will be sensitive to the GEM's inner hole diameters. By using MATLAB to analyze the images and convert pixel counts into distances, the pitch and the inner and outer hole diameters can be determined.
 
\section{Electrical Performance}\label{sec:electrical}
\begin{figure}[!t]
\centering
\includegraphics[width=\columnwidth]{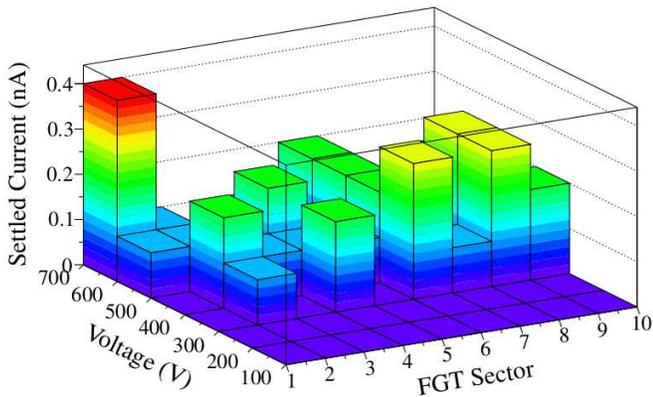}
\caption{Measured leakage current as a function of voltage and sector (1-9) for a single-mask 40$\times$40 cm$^2$ GEM foil. The measured current is accurate to within about 0.5 nA.}
\label{fig:current}
\end{figure}

The electrical performance of the GEM foil is determined through its leakage current measurement as described in section~\ref{sec:electric-ana}. All of the measured GEM foils from Tech-Etch (24 10$\times$10 and 3 40$\times$40 cm$^2$ foils) were found to have excellent electrical properties. A typical leakage current of less than 1 nA was consistently measured for all foils. These results were independently checked by Tech-Etch, who found similar results from measurements taken prior to shipping the GEM foils to Temple University. It is believed that the superb electrical performance of the GEMs is due to switching the polyimide layer from Kapton to Apical, which is known to have a much lower water absorption rate than Kapton~\cite{Apical,Kapton}. Earlier Tech-Etch GEMs which had used Kapton as the insulating material typically saw a leakage current on the order of 10 nA. The leakage current measurement from one of the 40$\times$40 GEM foils can be seen in fig.~\ref{fig:current}, where the current is plotted as a function of applied voltage and foil sector.   

\section{Geometrical Properties}\label{sec:geometrical}
The geometrical properties of the GEM foils were measured using the optical analysis setup described in section~\ref{sec:optical-ana}. These geometrical properties include the pitch between neighboring holes, the inner and outer hole diameters, and their uniformity across the entire active area of the GEM foil. This kind of optical analysis is critical in establishing a reliable commercial fabrication process, as inconsistencies in the geometrical parameters or their non-uniformity across a GEM foil will degrade its performance. As a result, an exhaustive optical analysis was carried out on the 10$\times$10 and 40$\times$40 cm$^2$ GEM foils.  

\subsection{10$\times$10 cm$^2$ Foils}
\begin{figure}[!t]
\centering
\includegraphics[width=\columnwidth]{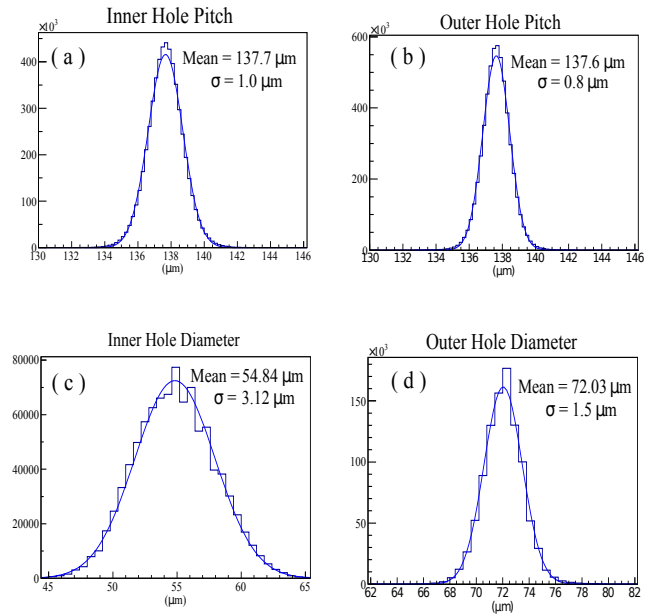}
\caption{Representative sample of the single-mask 10$\times$10 cm$^2$ GEM foils' geometrical distributions. (a): Pitch distribution measured using the back light. (b): Pitch distribution measured using the front light. (c): Inner hole diameter distribution. (d) Outer hole diameter distribution.}
\label{fig:10by10-distribution}
\end{figure}

The optical analysis of the last manufacturing lot of the 10$\times$10 cm$^2$ foils is presented here. Each of the six foils in this lot were found to have similar geometrical properties. Figure~\ref{fig:10by10-distribution} shows the pitch, and hole diameter distributions which were typical of all six foils. All of the distributions showed a Gaussian behavior, with the pitch having the narrowest distribution ($\sigma \sim 1\mu m$). The pitch measurement was found to be consistent at about 138 $\mu m$ when using the inner or outer hole diameters to determine its value. The inner hole diameter was found to have a wider distribution than the outer hole diameter, which is due to the inner hole etching process being more sensitive to the etching time than the copper layer etching process.

\begin{figure}[!t]
\centering
\includegraphics[width=\columnwidth]{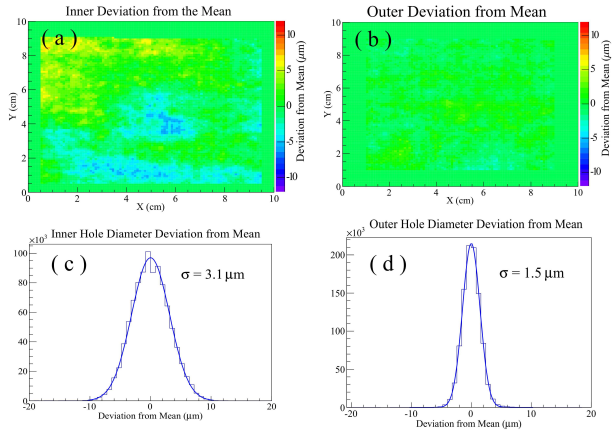}
\caption{Representative sample of the single mask 10$\times$10 cm$^2$ GEM foils' hole diameter uniformity. (a): Inner hole diameter uniformity. (b): Outer hole diameter uniformity. (c): Inner hole diameter deviation from mean. (d): Outer hole diameter deviation from mean.}
\label{fig:10by10-uniformity}
\end{figure}

The uniformity of the inner and outer hole diameters across the GEM foil also play a key role in determining the quality of the GEM foil. Large non-uniformities in the hole diameters could lead to a loss in the gain that is achievable with the foil. The inner and outer hole deviations were found to be small, about $\pm$ 6 $\mu m$ (fig.~\ref{fig:10by10-uniformity}(a-b)), with the larger deviations corresponding to the inner hole diameters (fig.~\ref{fig:10by10-uniformity}(c-d)). Simple model calculations have shown that such a deviation would not significantly affect the hit position or resolution of a potential tracking detector.

If we include the results for all six GEM foils into our analysis, we find that the pitch is nearly constant at about 138 $\mu m$ over all of the foils, and the mean inner (outer) hole diameter across all six foils is $\sim$ 58.47 $\mu m$ (71.58 $\mu m$). These results are consistent with other GEM foils which were produced via the double-mask process~\cite{Becker:2006}. Tech-Etch has also measured the inner and outer hole diameters for their single-mask 10$\times$10 cm$^2$ foils, however unlike Temple University which includes all of the holes on the GEM foil, Tech-Etch's measurements only included nine holes. Nonetheless, Tech-Etch's measurements of the inner and outer hole diameters are consistent with those found by Temple University.

\begin{figure}[!t]
\centering
\includegraphics[width=\columnwidth]{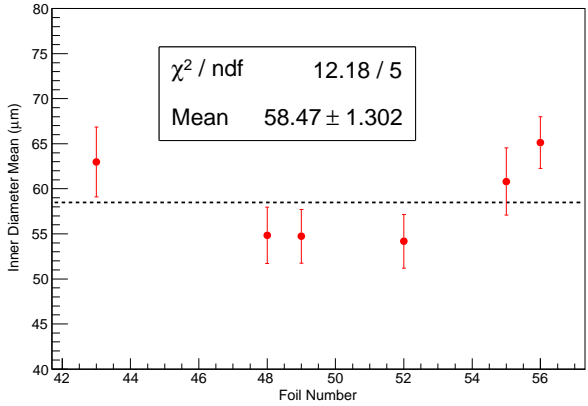}
\caption{Average inner hole diameter across last batch of Tech-Etch 10$\times$10 cm$^2$ single-mask GEM foils. The error bars represent the sigma of a Gaussian fit to that particular distribution.}
\label{fig:10by10-id}
\end{figure}

\begin{figure}[!t]
\centering
\includegraphics[width=\columnwidth]{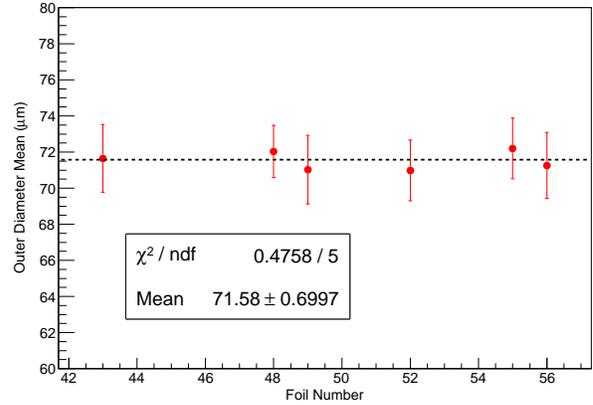}
\caption{Average outer hole diameter across last batch of Tech-Etch 10$\times$10 cm$^2$ single-mask GEM foils. The error bars represent the sigma of a Gaussian fit to that particular distribution.}
\label{fig:10by10-od}
\end{figure}

\subsection{40$\times$40 cm$^2$ Foils}
\begin{figure}[!t]
\centering
\includegraphics[width=0.5\columnwidth]{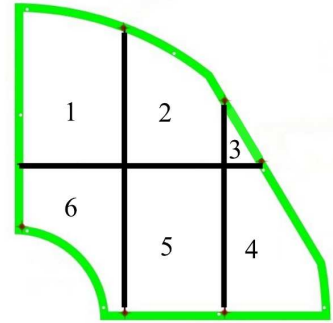}
\caption{The division of a 40$\times$40 cm$^2$ GEM foil into six CCD scan regions (1-6).}
\label{fig:fgt-split}
\end{figure}

The optical analysis of the larger 40$\times$40 cm$^2$ followed the same procedure used to measure the geometrical properties of the 10$\times$10 cm$^2$ foils. However, whereas all of 10$\times$10 cm$^2$ foil images were taken with one CCD scan, the 40$\times$40 cm$^2$ foils needed to be divided into six CCD scans due to the translational limitation of our 2D stage. Figure~\ref{fig:fgt-split} shows how the 40$\times$40 cm$^2$ foils were divided in order to scan the entire active area of the foil. 

\begin{figure}[!t]
\centering
\includegraphics[width=3.5in]{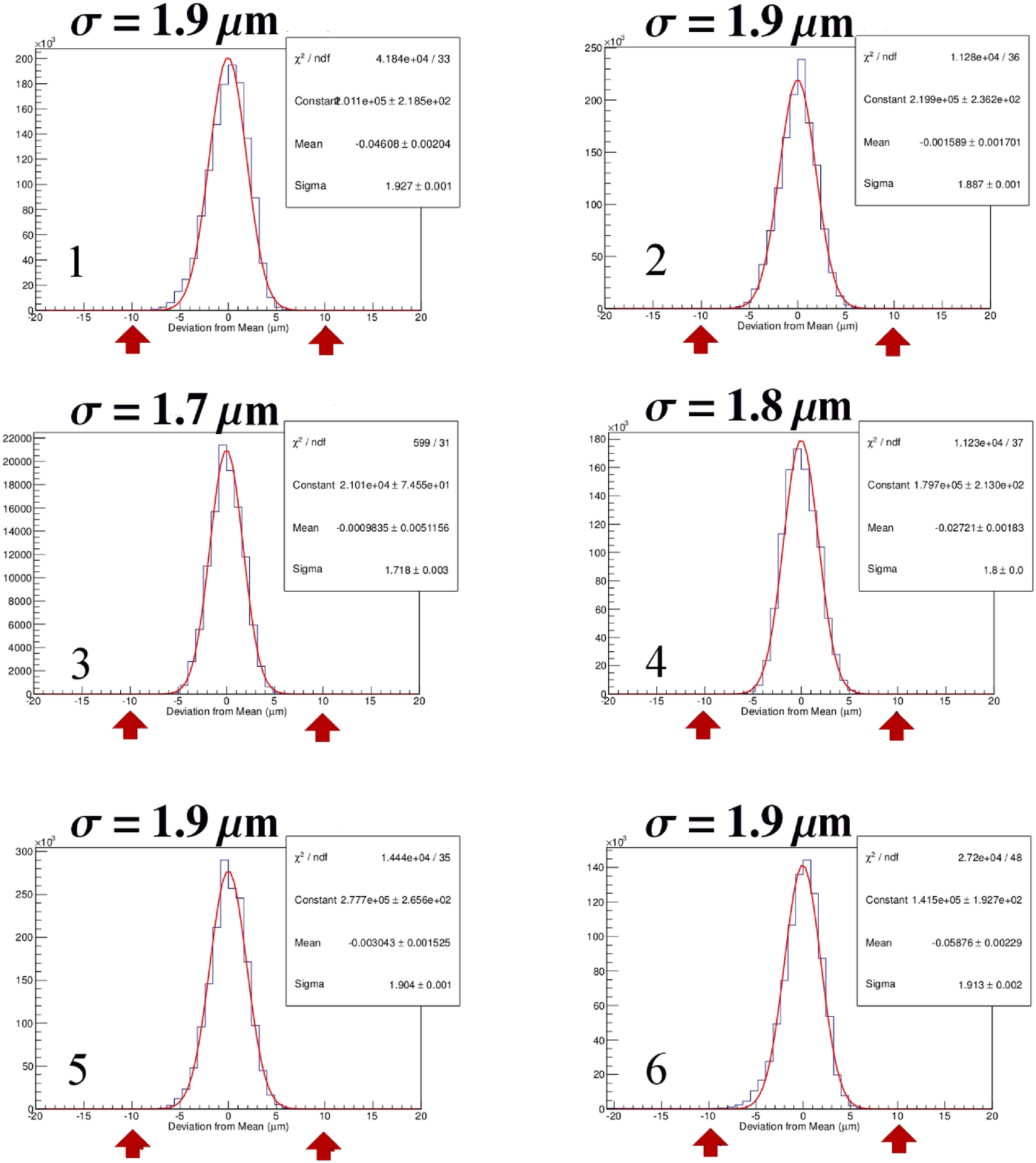}
\caption{Inner hole diameter deviation from mean for CCD scan regions 1-6. The red arrows mark the $\pm$10 $\mu m$ position.}
\label{fig:fgt-uniformity}
\end{figure}

Similar distributions (pitch, inner, and outer hole diameters) were measured with the 40$\times$40 cm$^2$ foils as were found in the 10$\times$10 cm$^2$ foils. Many of the same geometrical behaviors found in the 10$\times$10 cm$^2$ were also seen in the larger foils. In particular the pitch displayed the narrowest distribution and the inner hole diameters showed a larger deviation from the mean than the outer hole diameters. Also like the 10$\times$10 cm$^2$ foils, the hole diameters were found to have excellent uniformity across the 40$\times$40 cm$^2$ foils, where deviations were found to be smaller $\pm10 \mu m$, as shown in fig.~\ref{fig:fgt-uniformity}. The inner (outer) hole diameter deviation distribution widths generally ranged from $\sigma$ = 1.7 $\rightarrow$ 3.0 $\mu m$ ($\sigma$ = 1.1 $\rightarrow$ 1.8 $\mu m$).      

\begin{figure}[!t]
\centering
\includegraphics[width=\columnwidth]{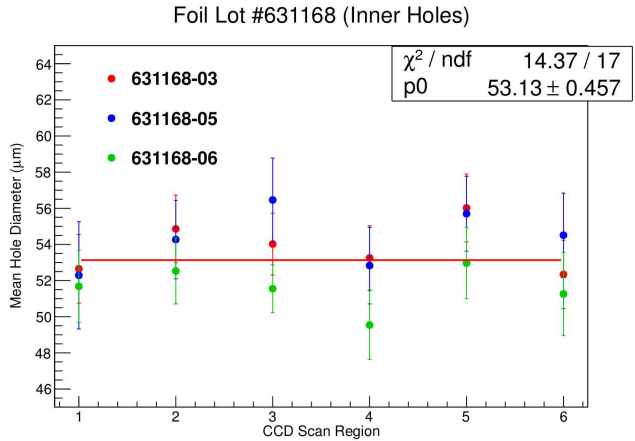}
\caption{Average inner hole diameters for each CCD scan region for all three 40$\times$40 cm$^2$ foils. The error bars represent the sigma of a Gaussian fit to that particular distribution. A constant line is fit to get the mean inner hole diameter across all of the foils.}
\label{fig:fgt-id}
\end{figure}

\begin{figure}[!t]
\centering
\includegraphics[width=\columnwidth]{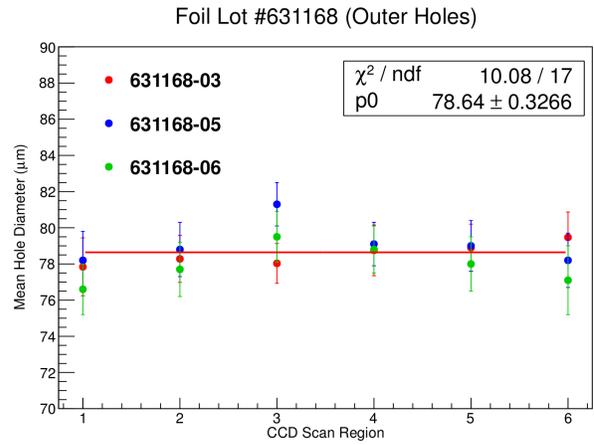}
\caption{Average outer hole diameters for each CCD scan region for all three 40$\times$40 cm$^2$ foils. The error bars represent the sigma of a Gaussian fit to that particular distribution. A constant line is fit to get the mean outer hole diameter across all of the foils.}
\label{fig:fgt-od}
\end{figure}

Considering all three of the 40$\times$40 cm$^2$ foils, we measured a near constant pitch of about 138 $\mu m$ in each CCD scan region across all foils. The average inner (outer) hole diameters were found to be consistent over all CCD scan regions across all three foils, as shown in fig.~\ref{fig:fgt-id} (fig.~\ref{fig:fgt-od}). The mean inner (outer) hole diameter across all three foils was measured to be 53.13 $\mu m$ (78.64 $\mu m$), which are similar to the double-mask GEM foil values found in ref.~\cite{Becker:2006}.   

\section{Summary}
Tech-Etch has successfully produced single-mask GEM foils of 10$\times$10 and 40$\times$40 cm$^2$, establishing for the first time the commercialization of GEM technology. All of the produced foils displayed outstanding electrical quality, with typical leakage currents repeatedly measured below about 1 nA. Both foil sizes were found to have similar geometric properties and behaviors, with the pitch remaining constant across all foils at about 138 $\mu m$, and the inner (outer) hole diameter being measured on the order of 50$\mu m$ (70 $\mu m$), which are in line with previously measured double-mask GEM foils. Excellent inner and outer hole diameter uniformity (less than $\pm10 \mu m$) was found across the entire active area of the 10$\times$10 and 40$\times$40 foils. Building on the success of these GEM foils, Tech-Etch is now in the beginning stages of looking into upgrading their production facility in order to produce GEM foils on the order of 1 m long.   
\vfill
\section*{Acknowledgment}
We would like to thank David Crary, Kerry Kearney, and Matthew Campbell (Tech-Etch Inc.), as well as M.~Hohlmann (FIT), R.~Majka (Yale), and especially R.~De~Oliveira (CERN) for their useful discussions, guidance, and expertise which has lead to the successful commercialization of GEM technology.


%


\end{document}